\documentclass[aps,prab,twocolumn,amsmath,amssymb,groupedaddress]{revtex4-2}

\usepackage{siunitx}
\usepackage{placeins}
\usepackage{float}
\usepackage[colorlinks=true]{hyperref}
\hypersetup{ 
    citecolor=black,%
    filecolor=black,%
    linkcolor=black,%
    urlcolor=black 
}
\usepackage{graphicx}
\usepackage{dcolumn}
\usepackage{bm}
\usepackage{color}

\newcommand*\diff{\mathop{}\!\mathrm{d}}

\begin{document}

\title{Optimization of the Resolution of a Streaking Setup}

\author{F.~Mayet}
\email{frank.mayet@desy.de}
\affiliation{Deutsches Elektronen-Synchrotron DESY, Notkestr. 85, 22607 Hamburg, Germany}
\author{K.~Floettmann}
\email{klaus.floettmann@desy.de}
\affiliation{Deutsches Elektronen-Synchrotron DESY, Notkestr. 85, 22607 Hamburg, Germany}

\date{\today}

\begin{abstract}
In low $\gamma$ regimes the inevitable induced energy spread in combination with longitudinal dispersion during the drift through a transverse deflecting structure (TDS) can alter the phase space of the particle bunch to be characterized. The consequential bunch lengthening can render measuring ultra-short bunches impossible, even for high available streaking powers and hence theoretical resolution. In this work the resulting effective resolution of a TDS is estimated analytically and the optimal streaking voltage is determined. The findings are subsequently benchmarked using matrix-based particle tracking. Furthermore, additional transverse, as well as other resolution limiting effects of a realistic streaking setup are discussed. A potential mitigation scheme using an additional downstream focusing element is evaluated.
\end{abstract}

\maketitle

\section{Introduction}
Many applications of particle accelerators demand for high-brightness beams. This is for example true for free-electron lasers \cite{RevModPhys.88.015006}, ultra-fast electron diffraction \cite{doi:10.1098/rsta.2005.1735}, and medical applications in the so-called eFLASH regime \cite{doi:10.1126/scitranslmed.3008973}. These kinds of beams are furthermore a requirement, or even an intrinsic feature of novel compact accelerator concepts (e.g. \cite{England:2014bf,KARTNER201624,Giorgianni:2016ds,Nghiem:437316}). The most used definition of brightness is \cite{doi:10.1063/1.860076}
\begin{equation}
	B = \frac{\eta I}{\pi^2 \varepsilon _x \varepsilon _y},
\end{equation}
where $\eta$ is a form factor, $\varepsilon_{x,y}$ is the transverse emittance, and $I$ is the beam peak current, which is a function of the beam current profile. In order to maximize $B$ it is hence necessary to be able to characterize the longitudinal properties of the bunch to high precision. One way to measure the bunch length, and ideally to fully characterize the shape of the current profile, is to employ a so-called transverse deflecting structure (TDS) \cite{osti_784939}. The electromagnetic fields inside of such a structure excert a transverse force on the bunch, which depends on the longitudinal intra-bunch coordinate. This way, the longitudinal plane can be mapped to the transverse plane and evaluated using a standard transverse beam profile monitor. The smallest detectable feature within the current profile depends on a number of parameters. These are related to the particle beam itself, the beam optics, as well as the electromagnetic field configuration and so called streaking power of the TDS. In the following discussion an ultimate limit of the achievable effective resolution of a TDS is derived analytically and benchmarked against simulations. This effective resolution takes the evolution of the beam phase space along the finite length of the TDS into account, which is especially important in low $\gamma$ regimes. The potential requirement to reduce the streaking power to improve the temporal resolution has been noticed already experimentally (e.g. \cite{PhysRevLett.118.154802}). Furthermore, recent plans for experiments at the ARES linac at DESY \cite{instruments5030028} require the characterization of sub-fs microbunches with a mean energy of $\sim$\SI{50}{\mega e\volt} \cite{MAYET2018213,Mayet:426707}. Simulations based on the ARES setup, which includes two PolariX X-band TDS \cite{PhysRevAccelBeams.23.112001} show that a reduction in streaking voltage below the design value of \SI{20}{\mega\volt} is required to achieve best possible resolution. Following these observations, in this work, we quantify the effective resolution of a TDS setup and find the optimal streaking voltage.
\section{Effective Resolution} \label{sec:effective_resolution}
A transverse deflecting structure, due to the Panofsky-Wenzel Theorem \cite{doi:10.1063/1.1715427}, induces both correlated and uncorrelated energy spread. The induced uncorrelated rms-spread of the relative energy deviation $\delta = \Delta \gamma / \gamma$ is given by
\begin{equation}
	\Delta\sigma _\delta = \frac{eVk}{pc}\cdot \sigma _y(z_0),
\end{equation}
where $z_0$ is the longitudinal center of the TDS, $k = 2 \pi f / c$ the wavenumber of the TDS field of frequency $f$, $V$ the integrated streaking voltage along the TDS, $p$ the mean momentum of the electron beam, $\sigma _y$ the transverse rms beam size, and $c$ the speed of light in vacuum. In the following $\sigma_y(z_0) \equiv \sigma_0$ for brevity. We also drop the subscript $y$, which denotes the streaking direction. Furthermore, all transverse properties refer to the streaking plane and a subscript $0$ to the longitudinal center of the TDS. Definition of $p\cdot c/e = \tilde p$ then yields the simplified form of the minimal induced uncorrelated relative energy spread
\begin{equation}
 	\Delta\sigma _\delta = \frac{Vk}{\tilde p} \cdot \sigma _0.
 	\label{eq:induced_sigma_delta}
 \end{equation}
It should be noted that this is the integrated effect, meaning that in principle 
\begin{equation}
	\Delta\sigma _\delta(z) = \frac{k}{\tilde p} \cdot \int _0^z G(z^\prime) \sigma (z^\prime) \diff z^\prime,
\end{equation}
where $G(z)$ is the streaking gradient and $\sigma(z)$ is the transverse beam envelope. In the following we assume a more or less collimated beam,  as well as a constant streaking gradient along the TDS, i.e. $G(z) \equiv G = \text{const.}$. 

In absence of dispersive effects and based on the unperturbed optics, the smallest detectable longitudinal feature is given by \cite{Behrens:DIPAC09}
\begin{equation}
	\sigma_{z,\text{min}} = \sqrt{\frac{\varepsilon}{\beta_0}} \cdot \frac{1}{|\sin (\Delta \Psi) \cdot \cos (\phi)|} \cdot \frac{\lambda p c}{2 \pi eV},
\end{equation}
where $\lambda = 2 \pi / k$, $\Delta \Psi$ is the betatron phase advance between TDS center and detector and $\phi$ is the phase of the TDS field. $\varepsilon$ and $\beta$ are the geometric emittance and the beta-function respectively, as usual. For now, we only consider the ideal zero-crossing case, where $\phi = 0$ and $\Delta \Psi = \pi / 2$ (see the Appendix for examples of how to achieve this optimal phase advance). Using $\sigma_0 = \sqrt{\varepsilon \beta _0}$ yields
\begin{equation}
	\sigma_{z,\text{min}} = \frac{\varepsilon}{\sigma_0} \cdot \frac{\tilde p}{Vk}.
	\label{eq:tds_resolution}
\end{equation}
The longitudinal displacement of a particle due to linear longitudinal dispersion $R_{56}$ and non-zero energy deviation $\Delta \gamma$ is given by
\begin{equation}
	\Delta z = R_{56} \delta,
	\label{eq:sig_z_dis}
\end{equation}
where again $\delta = \Delta \gamma / \gamma$. The longitudinal dispersion of a drift section is given by
\begin{equation}
	R_{56,\text{d}} = -\frac{d}{\beta^2\gamma^2},
	\label{eq:r56_drift}
\end{equation}
where $d$ is the drift distance. Here, $\beta = v/c$. As mentioned before, the TDS-induced $\Delta\sigma _\delta$ is an integrated effect. It is often assumed that the bunch length at the center of the TDS is reconstructed. Considering the aforementioned longitudinal dispersion, as well as the induced energy spread in the TDS, it is better to state that the \emph{average bunch length along the structure} is evaluated. In order to determine this average bunch length, it is hence required to average the bunch lengthening along the structure. Integration of Eq.~\ref{eq:sig_z_dis} up to $L$, where $L$ is the length of the TDS, yields the bunch lengthening as
\begin{equation}
	\begin{split}
	\Delta\sigma_{z,\text{dis}} &= \int _0 ^{L} \frac{G k}{\tilde p} \sigma_0 \cdot \frac{z}{\beta^2\gamma^2} \diff z \\
	&= \frac{V k L}{2} \cdot \frac{\sigma_0}{\tilde p \beta^2\gamma^2},
	\end{split}
	\label{eq:sigma_dis}
\end{equation}
where $V = \int _0 ^L G dz = GL$ was used (cf. Eq.~19 in \cite{PhysRevSTAB.17.024001}). The average bunch length change along the structure is then given by
\begin{equation}
	\begin{split}
	\overline{\Delta\sigma}_{z,\text{dis}} &= \frac{1}{L} \cdot \int _0 ^{L}  \int_0 ^z \frac{G k}{\tilde p} \sigma_0 \cdot \frac{z^\prime}{\beta^2\gamma^2} \diff z^\prime \diff z \\
	&= \frac{V k L}{6} \cdot \frac{\sigma_0}{\tilde p \beta^2\gamma^2}.
	\end{split}
	\label{eq:avg_sigma_dis}
\end{equation}

In an experiment, the measured beam size at a screen downstream of the TDS is given by
\begin{equation}
	\sigma_\text{scr}^2 = \sigma_\text{off}^2 + (S\cdot \sigma_z)^2,
\end{equation}
where $\sigma_\text{off}$ is the beam size in case the TDS is switched off. $S$ is the so-called shear parameter, which relates the screen coordinate system with the longitudinal plane of the electron bunch. The shear parameter can be obtained experimentally by performing a phase scan of the TDS, while recording the induced beam centroid offset on the screen. The bunch length can then be measured as 
\begin{equation}
	\sigma_{z,\text{mea}} = \frac{\sigma_\text{scr}}{S}.
	\label{eq:sig_z_meas}
\end{equation}
Inserting the definition of $\sigma_\text{scr}$ yields
\begin{equation}
	\sigma_{z,\text{mea}}^2 = \frac{\sigma_\text{off}^2}{S^2} + \sigma_z^2.
\end{equation}
From the discussion above, we know that indeed $\sigma_z \neq \sigma_{z,0}$, where $\sigma_{z,0}$ is the bunch length at the TDS center in case the TDS is switched off. Instead we have to set $\sigma_z^2 = \sigma_{z,0}^2 + \overline{\Delta\sigma}_{z,\text{dis}}^2$. The measured bunch length is hence given by
\begin{equation}
		\sigma_{z,\text{mea}}^2 = \sigma_{z,\text{min}}^2 + \sigma_{z,0}^2 + \overline{\Delta\sigma}_{z,\text{dis}}^2,
		\label{eq:sig_z_meas_mod}
\end{equation}
where the definition of the TDS resolution $R = \sigma_\text{off}/S \equiv \sigma_{z,\text{min}}$ was used. Clearly we require $\sigma_{z,\text{min}}^2 + \overline{\Delta\sigma}_{z,\text{dis}}^2 \ll \sigma_{z,0}^2$, in order to resolve $\sigma_{z,0}$. 

Based on Eq.~\ref{eq:sig_z_meas_mod} one could argue that it should be possible to retrieve $\sigma_{z,0}$ by quadratically subtracting the contributions from the theoretical resolution and the average bunch lengthening along the structure. This is indeed the case, if only the rms bunch length is of interest. In fact, depending on the screen resolution, one could even use a very low streaking voltage, such that the streaking is just detectable. In practice, however, the goal is often to characterize the full longitudinal profile of the bunch. This means that potentially fine sub-structures need to be resolved. Subtraction of a constant does not help here, as the measured screen image must be understood as the convolution of the unperturbed transverse beam shape, the broadening due to bunch lengthening, and the actual longitudinal bunch shape via streaking. Mathematically this means that
\begin{equation}
	f_\text{mea}(y) = f_{z,\text{min}}(y) \ast f_{z,\text{dis}}(y) \ast f_{z,0}(y),
\end{equation}
where the subscripts are chosen according to Eq.~\ref{eq:sig_z_meas_mod}. Deconvolution is not generally possible here. It can hence be concluded that in case the longitudinal profile of the bunch is of interest, an optimal streaking voltage needs to be found, balancing streaking power and induced bunch lengthening. Without loss of generality, this can be done based on the rms values used so far.

Inserting the respective definitions established above and defining  
\begin{equation}
	\sigma_{z,\text{eff}} = \sqrt{\sigma_{z,\text{min}}^2 + \overline{\Delta\sigma}_{z,\text{dis}}^2}
\end{equation}
as the effective resolution of the TDS yields
\begin{equation}
	\begin{split}
		 \sigma_{z,\text{eff}}^2 &= \frac{\varepsilon_\text{n}^2\tilde p^2}{\beta^2\gamma^2V^2k^2\sigma_0^2} + \frac{V^2k^2\sigma_0^2L^2}{36 \tilde p^2 \beta^4\gamma^4}\\
		&= \frac{a^2}{V^2} + b^2 V^2,
	\end{split}
\end{equation}
where $a = \varepsilon_\text{n}\tilde p/(\beta\gamma k\sigma_0)$, $b = k \sigma_0 L /(6 \tilde p \beta^2\gamma^2)$ and $\varepsilon_\text{n} = \beta \gamma \cdot \varepsilon$. Using this, it is possible to find the optimal streaking voltage via
\begin{equation}
	\frac{\diff}{\diff V} \left(\frac{a^2}{V^2} + b^2 V^2 \right) = 2b^2V - \frac{2a^2}{V^3}
\end{equation}
and subsequently
\begin{equation}
		2b^2V = \frac{2a^2}{V^3} \Leftrightarrow V = \sqrt[4]{\frac{a^2}{b^2}}.
\end{equation}
The optimal streaking voltage is hence given by
\begin{equation}
	V_\text{opt} = \sqrt{\frac{6\varepsilon_\text{n}\beta\gamma}{L}} \cdot \frac{\tilde p}{k \sigma_0}.
\end{equation}
If $V < V_\text{opt}$, the resolution is limited by the streaking power of the TDS. On the other hand, if $V > V_\text{opt}$, the resolution is limited by the bunch lengthening inside of the TDS. From this, the optimal effective resolution follows as
\begin{equation}
 	\left. \sigma_{z,\text{eff}}\right|_{V=V_\text{opt}} = \sqrt{\frac{\varepsilon _\text{n}L}{3 \beta^3\gamma^3}}.
\end{equation}
It is important to note that in reality the maximum achievable streaking voltage might be smaller than $V_\text{opt}$. In this case, the resolution is limited by the streaking power. 

In case the beam transport from TDS to the screen is just a drift of length $d$, the TDS resolution can be calculated as
\begin{equation}
	\sigma_{z,\text{min,d}} = \frac{\sigma_\text{off}}{d} \cdot \frac{\tilde p}{Vk}.
\end{equation}
The optimal streaking voltage is then given by
\begin{equation}
	V_\text{opt,d} = \sqrt{\frac{6\sigma_\text{off}}{d\sigma_0 L}} \cdot \frac{\tilde p \beta \gamma}{k}
\end{equation}
and
\begin{equation}
	\left. \sigma_{z,\text{eff}}\right|_{V=V_\text{opt,d}} = \sqrt{\frac{\sigma_\text{off}\sigma_0 L}{3 \beta^2\gamma^2d}}.
\end{equation}

Using these findings, it is possible to determine optimal TDS working points for given beam parameters.

In \cite{PhysRevSTAB.17.024001} the authors examine the relative contribution of the defocusing effect onto the beam by calculating the ratio
\begin{equation}
	\frac{p^\text{def}}{p^\text{shear}} = \frac{\varepsilon}{R^2 \beta \gamma^2}\cdot \frac{L}{6}.
\end{equation}
Inserting $V_\text{opt}$ reveals the interesting observation that
\begin{equation}
	\left. \frac{p^\text{def}}{p^\text{shear}} \right|_{V=V_\text{opt}} = \beta,
\end{equation}
or in other words, the optimal streaking voltage is reached when the contribution of the defocusing effect and the streaking is balanced, scaled by $\beta$.
\section{Particle Tracking}
In this section the concept of an effective resolution is verified through matrix-based particle tracking. The symplectic transfer matrix of a streaking cavity acting on the phase space vector $(y,y^\prime,\zeta,\delta)^T$ is given by \cite{PhysRevSTAB.5.084001}
\begin{equation}
	\mathbf{M}_\text{TDS} = 
	\begin{pmatrix}
		1 & L & KL/2 & 0\\
		0 & 1 & K & 0 \\
		0 & 0 & 1 & 0 \\
		K & KL/2 & K^2L/6 & 1
	\end{pmatrix},
\label{eq:TDS_matrix}
\end{equation}
where $K = Vk / \tilde p$ and $y$ is the streaking direction. $\zeta$ is the longitudinal intra-bunch coordinate with respect to the bunch center and $\delta = \Delta \gamma / \gamma_0$. In this study, only a drift between the TDS and the screen is considered. The transfer matrix for a simple drift, including linear longitudinal dispersion $R_{56}$ is given by
\begin{equation}
	\mathbf{M}_\text{D}(s) = 
	\begin{pmatrix}
		1 & s & 0 & 0\\
		0 & 1 & 0 & 0\\
		0 & 0 & 1 & R_{56}(s)\\
		0 & 0 & 0 & 1
	\end{pmatrix},
\end{equation}
where $s$ is the drift distance. Notably Eq.~\ref{eq:TDS_matrix} does not include any longitudinal dispersion due to the drift distance $L$. In order to simulate the dynamics, which cause the bunch lengthing described above, Eq.~\ref{eq:TDS_matrix} needs to be modified, such that
\begin{equation}
	\mathbf{\tilde M}_\text{TDS}(L) = 
	\begin{pmatrix}
		1 & L & KL/2 & 0\\
		0 & 1 & K & 0 \\
		0 & 0 & 1 & R_{56}(L) \\
		K & KL/2 & K^2L/6 & 1
	\end{pmatrix}.
\label{eq:TDS_matrix_mod}
\end{equation}
Furthermore, to capture the evolution of the longitudinal phase space inside of the TDS, $\mathbf{\tilde M}_\text{TDS}$ is split into $N$ parts. Hence for each part $l = L / N$. The particle transport is now given by
\begin{equation}
 	\begin{pmatrix}
 		y \\ y^\prime \\ \zeta \\ \delta
 	\end{pmatrix}
 	= \mathbf{M}_\text{D}(d_\text{s}) \cdot  \prod ^N \mathbf{\tilde M}_\text{TDS}(l) \cdot
 	\begin{pmatrix}
 		y_0 \\ y^\prime_0 \\ \zeta_0 \\ \delta_0
 	\end{pmatrix},
\end{equation}
where $d_\text{s}$ is the drift from the end of the TDS to the screen. We note that in order to fully capture the dynamics within the TDS, the minimal value of $N$ has to be determined via a convergence study. In this study $N = 1000$ was found to be sufficient. The reconstruction of the bunch length can in this configuration be calculated using Eq.~\ref{eq:sig_z_meas} and 
\begin{equation}
	S_\text{d} = \frac{dVk}{\tilde p},
\end{equation}
where in our case $d = L/2 + d_\text{s}$. 

\begin{figure}[htbp]
  \centering
  \includegraphics[width=\columnwidth]{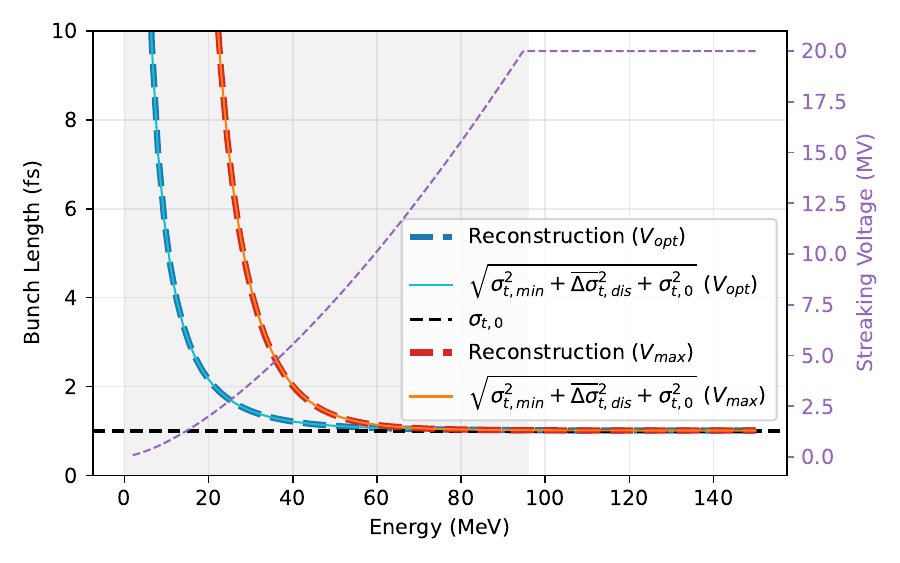}
  \caption{Plot of the effective resolution vs. beam energy for a case where $V = V_\text{opt}$ (blue, cyan) and a case where is maximum available streaking voltage $V_\text{max} = \SI{20}{\mega\volt}$ is used (red, orange). $V_\text{opt}$ is restricted to be smaller than $V_\text{max}$. In this example $\varepsilon _\text{n} = \SI{80}{\nano\meter}$, $\sigma_0 = \SI{200}{\micro\meter}$, $L = \SI{0.8}{\meter}$ and $f = \SI{11.9952}{\giga\hertz}$ (X-band). The initial bunch length is \SI{1}{\femto\second} and $\sigma_{\delta,0} = 0$ to study the fundamental resolution limit. The shaded area corresponds to the bunch lengthening limited regime. The blue and red dashed lines correspond to the simulated data.}
  \label{fig:POLARIX_example_tracking}
\end{figure}

As an example we use the PolariX X-band TDS and run the simulation for different beam energies. The scan is performed twice with 100,000 particles per bunch. First, for each energy, the optimal streaking voltage $V_\text{opt,d}$ is calculated and used. Then the scan is repeated with a fixed voltage $V_\text{max}$. The result is shown in Fig.~\ref{fig:POLARIX_example_tracking}. In each case both the reconstructed bunch length (dashed line), as well as the expected measured bunch length according to the analytical Eq.~\ref{eq:sig_z_meas_mod} (thin solid line) is plotted. It can be seen that the analytical estimate and the simulation data agree very well. The results show indeed that in the bunch lengthening limited regime the streaking voltage must be reduced significantly, in order to optimize the effective resolution. At an exemplary beam energy of \SI{25}{\mega e\volt} it can be seen that the measured bunch length using $V = V_\text{opt} = \SI{2.8}{\mega \volt}$ is \SI{1.7}{\femto \second}, wheras using $V = V_\text{max} = \SI{20}{\mega \volt}$ it is \SI{7.1}{\femto \second}. Using $V_\text{opt}$ at this particular energy hence results in a $\sim 4\times$ better resolution.
\section{Transverse Effects}
In addition to the longitudinal effects discussed above, there are transverse effects, which can impair the achievable resolution of the system. Firstly, field nonlinearities can lead to emittance growth in the streaking plane, beyond the desired one, which is caused by the streaking action itself. This leads to an increased $\sigma_{z,\text{min}}$. Secondly, the bunch lengthening discussed above leads to defocusing in the streaking plane, which is independent of the streaking action \cite{PhysRevSTAB.17.024001}. In order to describe these two effects - in contrast to how Eq.~\ref{eq:tds_resolution} is derived - a full matrix description of the streaking based on the perturbed optics needs to be considered.
\subsection{Full matrix description - Perturbed optics}
In the following only the streaking plane is discussed. We consider an unstreaked beam with the initial optical functions $\beta_\text{u,0}$, $\alpha_\text{u,0}$ and $\gamma_\text{u,0}$, which are transformed with an optical transfer matrix $\mathbf{T}$ to the screen position, where the beta-function of the unstreaked beam is denoted as $\beta_\text{u,scr}$. Furthermore, the geometrical emittance of the unstreaked beam is denoted as $\varepsilon_\text{u}$.

The streaking leads to an increase of the geometrical emittance given by \cite{PhysRevSTAB.17.024001}
\begin{equation}
	\varepsilon^2_\text{s} = \varepsilon^2_\text{u} + \left(\frac{Vk}{\tilde p} \sigma_z \sigma_0\right)^2,
\end{equation}
where the subscript s refers to the streaked beam. The beam size $\sigma_0$ does not change due to the streaking, thus the beta-function has to change as
\begin{equation}
	\beta_\text{s,0} = \frac{\sigma_0^2}{\varepsilon_\text{s}} = \beta_\text{u,0}\frac{\varepsilon_\text{u}}{\varepsilon_\text{s}}.
\end{equation}
The correlated beam divergence, i.e. the term $\langle yy^\prime \rangle$ stays constant, where $\langle \rangle$ denotes the second central moment. Therefore
\begin{equation}
	\alpha_\text{s,0} = -\frac{\langle yy^\prime \rangle}{\varepsilon_\text{s}} = \alpha_\text{u,0}\frac{\varepsilon_\text{u}}{\varepsilon_\text{s}}
\end{equation}
and
\begin{equation}
	\gamma_\text{s,0} = \frac{1+\alpha^2_\text{s,0}}{\beta_\text{s,0}} = \frac{\frac{\varepsilon^2_\text{s}}{\varepsilon^2_\text{u}} + \alpha^2_\text{u,0}}{\beta_\text{u,0}}\cdot\frac{\varepsilon_\text{u}}{\varepsilon_\text{s}}.
\end{equation}

Denoting the elements of the single particle transfer matrix $\mathbf{M}$ between TDS and screen as usual with cosine- and sine-like trajectories $C, S, C^\prime$ and $S^\prime$, yields the mapping
\begin{equation}
	\begin{pmatrix}
 		y_\text{scr} \\ y^\prime_\text{scr}
 	\end{pmatrix} 
 	= \mathbf{M} \cdot
 	\begin{pmatrix}
 		y_\text{0} \\ y^\prime_\text{0}
 	\end{pmatrix}
 	=
 	\begin{pmatrix}
 		C  & S \\ C ^\prime & S^\prime
 	\end{pmatrix} \cdot
 	\begin{pmatrix}
 		y_\text{0} \\ y^\prime_\text{0}
 	\end{pmatrix}.
\end{equation}
The optical transfer matrix is given by
\begin{equation}
	\mathbf{T} =
	\begin{pmatrix}
		C^2 & -2SC & S^2 \\
		-CC^\prime & SC^\prime + S^\prime C & -SS^\prime \\
		C^{\prime 2} & -2S^\prime C^\prime & S^{\prime 2}
	\end{pmatrix}.
\end{equation}
The principal trajectories can be represented in terms of the undistorted optics as
\begin{equation}
	\begin{split}
		C &= \sqrt{\frac{\beta_\text{u,scr}}{\beta_\text{u,0}}} (\cos (\Delta \Psi) + \alpha_\text{u,0}\sin (\Delta \Psi)), \\
		S &= \sqrt{\beta_\text{u,scr}\beta_\text{0,scr}} \sin (\Delta \Psi).
	\end{split}
\label{eq:CandS}
\end{equation}
From the mapping
\begin{equation}
	\begin{pmatrix}
		\beta_\text{u,scr} \\ \alpha_\text{u,scr} \\ \gamma_\text{u,scr}
	\end{pmatrix}
	= \mathbf{T} \cdot
	\begin{pmatrix}
		\beta_\text{u,0} \\ \alpha_\text{u,0} \\ \gamma_\text{u,0}
	\end{pmatrix}
\end{equation}
we hence get
\begin{equation}
	\begin{split}
		\beta_\text{u,scr} &= C^2 \beta_\text{u,0} - 2SC\alpha_\text{u,0} + S^2\gamma_\text{u,0} \\
		&= C^2 \beta_\text{u,0} - 2SC\alpha_\text{u,0} + S^2\frac{1+\alpha^2_\text{u,0}}{\beta_\text{u,0}}
	\end{split}
\end{equation}
and subsequently for the streaked case
\begin{equation}
	\begin{split}
		\beta_\text{s,scr} &= C^2 \beta_\text{s,0} - 2SC\alpha_\text{s,0} + S^2\gamma_\text{s,0} \\
		&= \frac{\varepsilon_\text{u}}{\varepsilon_\text{s}} \left( C^2 \beta_\text{u,0} - 2SC\alpha_\text{u,0} + S^2\frac{\frac{\varepsilon^2_\text{s}}{\varepsilon^2_\text{u}} + \alpha^2_\text{u,0}}{\beta_\text{u,0}} \right).
	\end{split}
\end{equation}

The maximal resolution power is achieved when the ratio $\rho$ of the streaked to the unstreaked beam sizes on the screen is maximized as 
\begin{equation}
	\rho = \frac{\sigma_\text{s,scr}}{\sigma_\text{u,scr}} = \sqrt{\frac{\beta_\text{s,scr} \varepsilon_\text{s}}{\beta_\text{u,scr} \varepsilon_\text{u}}} \geq 1.
\end{equation}
Taking the square yields
\begin{equation}
	\rho^2 = \frac{C^2 \beta_\text{u,0} - 2SC\alpha_\text{u,0} + S^2\frac{\frac{\varepsilon^2_\text{s}}{\varepsilon^2_\text{u}} + \alpha^2_\text{u,0}}{\beta_\text{u,0}}}{C^2 \beta_\text{u,0} - 2SC\alpha_\text{u,0} + S^2\frac{1+\alpha^2_\text{u,0}}{\beta_\text{u,0}}}.
\end{equation}
Inserting Eq.~\ref{eq:CandS} leads to the simplified expression
\begin{equation}
	\rho^2 = \cos^2(\Delta \Psi) + \frac{\varepsilon_\text{s}^2}{\varepsilon_\text{u}^2} \sin^2(\Delta \Psi),
\end{equation}
which is maximized with $\Delta \Psi = \pi/2$. As in Sec.~\ref{sec:effective_resolution} we assume this ideal case. Then
\begin{equation}
		\left.\rho\right|_{\Delta\Psi=\frac{\pi}{2}} = \frac{\varepsilon_\text{s}}{\varepsilon_\text{u}} = \sqrt{1+\frac{\left( \frac{Vk}{\tilde p} \sigma_z \sigma_0\right)^2}{\varepsilon_\text{u}^2}}.
\label{eq:rho_ideal}
\end{equation}
We define the resolution $\sigma_\text{z,min}$ as the bunch length at which $\rho = \sqrt{2}$, as usual. Referring to Eq.~\ref{eq:rho_ideal}, this means that
\begin{equation}
 	\frac{\frac{Vk}{\tilde p} \sigma_{z,\text{min}} \sigma_0}{\varepsilon_\text{u}} = 1
 \end{equation} 
and subsequently
\begin{equation}
	\sigma_{z,\text{min}} = \frac{\varepsilon_\text{u}}{\sigma_0} \cdot \frac{\tilde p}{Vk}.
\end{equation}
Dropping the subscript u, this is exactly Eq.~\ref{eq:tds_resolution}, as expected. Equation~\ref{eq:rho_ideal} now allows the inclusion of the emittance growth due to nonlinear fields.
\subsection{Reduction of resolution due to field nonlinearities}
The emittance growth due to field nonlinearities scales as \cite{PhysRevSTAB.17.024001}
\begin{equation}
	\varepsilon_\text{nl} = \mathcal{C}_0 \frac{V}{\tilde p} \sigma_0^4,
	\label{eq:non_linear_emittance}
\end{equation}
where $\mathcal{C}_0$ is a constant describing the strength of the distortion, which has to be determined for the individual TDS. Equation~\ref{eq:non_linear_emittance} can now be introduced in Eq.~\ref{eq:rho_ideal}, which yields
\begin{equation}
	\begin{split}
	\left.\tilde\rho\right|_{\Delta\Psi=\frac{\pi}{2}} &= \sqrt{\frac{\varepsilon_\text{s}^2 + \varepsilon_\text{nl}^2}{\varepsilon_\text{u}^2+ \varepsilon_\text{nl}^2}} \\
	&= \sqrt{1+\frac{\left( \frac{Vk}{\tilde p} \sigma_z \sigma_0\right)^2 }{\varepsilon_\text{u}^2 + \left( \mathcal{C}_0 \frac{V}{\tilde p} \sigma_0^4 \right)^2}}.
	\end{split}
\end{equation}
Note that the nonlinearities also act on the unstreaked transverse part of the beam. By using the same definition of the resolution as above, one arrives at
\begin{equation}
	\frac{\left( \frac{Vk}{\tilde p} \sigma_z \sigma_0\right)^2}{\varepsilon_\text{u}^2 + \left( \mathcal{C}_0 \frac{V}{\tilde p} \sigma_0^4 \right)^2} = 1
\end{equation}
and subsequently
\begin{equation}
	\tilde \sigma_{z,\text{min}} = \frac{1}{\sigma_0}\cdot \frac{\tilde p}{Vk}\cdot \sqrt{\varepsilon_\text{u}^2 + \left( \mathcal{C}_0 \frac{V}{\tilde p}\sigma_0^4\right)^2}.
	\label{eq:sigma_z_min_tilde}
\end{equation}
From this we get
\begin{equation}
		\tilde \sigma_{z,\text{min}}^2 = \underbrace{\left(\frac{\varepsilon}{\sigma_0}\cdot \frac{\tilde p}{Vk} \right)^2}_{\sigma_{z,\text{min}}^2} + \underbrace{\left( \frac{\mathcal{C}_0 \sigma_0^3}{k}\right)^2}_{\sigma_{z,\text{min,nl}}^2},
		\label{eq:sigma_z_min_tilde_sq}
\end{equation}
where we have again dropped the subscript u. It can be seen that the nonlinear contributions degrade the theoretically achievable resolution, strongly depending on the unstreaked beam size $\sigma_0$. By differentiating Eq.~\ref{eq:sigma_z_min_tilde_sq} with respect to $\sigma_0$, the optimal unstreaked beam size, which minimizes $\tilde \sigma_{z,\text{min}}$, can be calculated as
\begin{equation}
	\sigma_{0,\text{opt}} = \frac{1}{\sqrt[8]{3}} \cdot \sqrt[4]{\frac{\varepsilon \tilde p}{V \mathcal{C}_0}}.
\end{equation}
Inserting $\sigma_{0,\text{opt}}$ into Eq.~\ref{eq:sigma_z_min_tilde} yields
\begin{equation}
	\begin{split}
		\tilde \sigma_{z,\text{min}}(\sigma_{0,\text{opt}}) &= \frac{1}{\sigma_{0,\text{opt}}}\cdot \frac{\tilde p}{Vk}\cdot \sqrt{\varepsilon^2 + \left( \mathcal{C}_0 \frac{V}{\tilde p} \cdot \frac{1}{\sqrt{3}} \frac{\varepsilon \tilde p}{V \mathcal{C}_0}\right)^2} \\
		&= \frac{1}{\sigma_{0,\text{opt}}}\cdot \frac{\tilde p}{Vk}\cdot \sqrt{\varepsilon^2 + \frac{\varepsilon^2}{3}} \\
		&= \sigma_{z,\text{min}}(\sigma_{0,\text{opt}}) \cdot \sqrt{1.\overline{3}},
	\end{split}
\end{equation}
or in other words, if nonlinearities are present, the best resolution can be achieved when the combination of streaking voltage and beam size leads to an emittance growth of $\sqrt{1.\overline{3}} \approx \SI{15}{\percent}$.

We note that the optimal streaking voltage $V_\text{opt}$, derived in Sec.~\ref{sec:effective_resolution}, does not change as $\sigma_{z,\text{min,nl}}$ does not depend on $V$.
\subsection{Defocusing due to bunch lengthening}
As described in \cite{PhysRevSTAB.17.024001}, the bunch lengthening inside of the TDS leads to defocusing in the streaking plane. The total induced transverse momentum reads as
\begin{equation}
	p_y = a(-\Delta s_0+by), 
	\label{eq:py_along_the_tds}	
\end{equation}
(cf. Eq.~21 in \cite{PhysRevSTAB.17.024001}). Here $a = \frac{eVk}{c}\cos (\phi_0)$, $b = \frac{eVk}{cp\beta\gamma^2} \cos (\phi_0) \cdot \frac{L}{6}$, and $\Delta s_0$ denotes the initial longitudinal position of the particle relative to the bunch center. It can be seen that, due to the bunch lengthening induced second term $aby$, a defocusing is introduced, which is independent from the streaking action. By considering that 
\begin{equation}
	\langle y p_y\rangle = \langle ya(-\Delta s_0 + by)\rangle = ab\cdot \langle y^2 \rangle \equiv ab \cdot \sigma_0^2
\end{equation}
we can calculate the variation of the alpha-function due to the defocusing as
\begin{equation}
	\Delta \alpha = -\frac{\langle yp_y \rangle}{p\varepsilon _\text{s}} = - \frac{ab\cdot \sigma_0^2}{p \varepsilon_\text{s}}.
\end{equation}
Inserting $a$ and $b$ yields
\begin{equation}
	\Delta \alpha = - \left( \frac{Vk}{\tilde p}\right)^2 \cdot \frac{L}{6\beta\gamma^2}\cdot \beta_{\text{s},0},
	\label{eq:defocusing_due_to_lengthening}
\end{equation}
where we use $\phi_0 = 0$ as above.

In order to counteract the induced $\Delta \alpha$, a focusing element downstream of the TDS can be employed. By assuming a thin lens element with
\begin{equation}
	\mathbf{M}_\text{TL} = 
	\begin{pmatrix}
		1 & 0\\
		-k_\text{f}s_\text{eff} & 1\\
	\end{pmatrix},
\end{equation}
where $k_\text{f}$ is the focusing strength and $s_\text{eff}$ is the effective length of the lens, the same $\Delta \Psi$ at the screen as without defocusing can be easily achieved. In fact, both $\Delta \Psi$ and $\beta _\text{s,scr}$ can be preserved. The alpha-function after the thin lens is given by
\begin{equation}
	\begin{split}
		\alpha &= -m_{11}m_{21} \beta_{\text{s},0} \\& \hspace{1cm}+ (m_{12}m_{21} + m_{22}m_{11})\alpha_{\text{s},0}\\ & \hspace{1cm}- m_{12}m_{22}\gamma_{\text{s},0}\\
		&= k_\text{f}s_\text{eff}\beta_{\text{s},0} + \alpha_{\text{s},0},
	\end{split}
\end{equation}
where $m_{ij}$ are the matrix elements of $\mathbf{M}_\text{TL}$. We now require that $\alpha \overset{!}{=} \alpha_{\text{s},0} - \Delta \alpha$. Hence,
\begin{equation}
	\begin{split}
		k_\text{f}s_\text{eff}\beta_{\text{s},0} + \alpha_{\text{s},0} &= \alpha_{\text{s},0} - \Delta \alpha\\
		\Leftrightarrow k_\text{f} &= -\frac{\Delta \alpha}{\beta_{\text{s},0}s_\text{eff}}.
	\end{split}
\end{equation}
Inserting Eq.~\ref{eq:defocusing_due_to_lengthening} yields
\begin{equation}
	k_\text{f} = \frac{L}{6\beta s_\text{eff}\gamma^2} \cdot \left(\frac{Vk}{\tilde p} \right)^2.
	\label{eq:kf_zero_drift}
\end{equation}
This result is only valid if there is no drift between the TDS and the lens. Otherwise, due to $\Delta \alpha$, the beta-function and subsequently $\Delta \Psi$ will change during the drift. In the finite drift case, it is not generally possible to preserve both $\Delta \Psi$ and the beta-function at the screen. The goal should be to preserve $\Delta \Psi$, or in simple words the image should be kept sharp. This in turn means that the final beta-function will be smaller than $\beta_\text{s,scr}$. The lens thus de-magnifies the image. This does not reduce the resolution of the system per-se, but might affect it indirectly via e.g. the pixel size of the detector. Since now also the streaking action is de-magnified, in the experiment, the streaking parameter $S$ needs to be downscaled. The required $k_\text{f}$ for a finite drift length $d_\text{l}$ between TDS and lens can be calculated by solving $\Delta \Psi (\alpha_{\text{s},0},k_\text{f}=0) - \Delta \Psi (\alpha_{\text{s},0}+\Delta \alpha,k_\text{f}>0) = 0$ for $k_\text{f}$ and hence
\begin{equation}
	\begin{split}
	k_\text{f} &= \frac{(d_\text{l}+d_\text{ls})^2 \Delta \alpha}{s_\text{eff}d_\text{ls}(d_\text{l}(d_\text{l}+d_\text{ls})\Delta \alpha - \beta_{\text{s},0}d_\text{ls})}\\
	&=\frac{(d_\text{l}+d_\text{ls})^2 k^2 L V^2}{s_\text{eff}d_\text{ls}(6 \beta d_\text{ls} \gamma^2 \tilde p^2 + d_\text{l}(d_\text{l}+d_\text{ls})k^2LV^2)} ,
	\end{split}
	\label{eq:kf_finite_drift}
\end{equation}
where $d_\text{ls}$ is the distance between lens and screen and thus $d_\text{l}+d_\text{ls} = d_\text{s}$. We note that setting $d_\text{l} = 0$ in Eq.\ref{eq:kf_finite_drift} yields Eq.~\ref{eq:kf_zero_drift} as expected. The de-magnification factor $\mathcal{M}$ is given by the ratio of $\beta_\text{s,scr}(\alpha_{\text{s},0}+\Delta \alpha,k_\text{f}>0)$ and $\beta_\text{s,scr}(\alpha_{\text{s},0},k_\text{f}=0)$ and thus
\begin{equation}
	\begin{split}
	\mathcal{M} &= \frac{\beta_\text{s,scr}(\alpha_{\text{s},0}+\Delta \alpha,k_\text{f}>0)}{\beta_\text{s,scr}(\alpha_{\text{s},0},k_\text{f}=0)}\\
	&= \frac{(6\beta d_\text{ls}\gamma^2\tilde p^2)^2}{(6\beta d_\text{ls}\gamma^2\tilde p^2 + d_\text{l}(d_\text{l}+d_\text{ls})k^2LV^2)^2}.
	\end{split}
\end{equation}
It can be seen that $\mathcal{M}(d_\text{l}=0) = 1$, as expected, and that $\mathcal{M} < 1, \forall d_\text{l} > 0$. This factor can be used to scale the shear parameter $S$ as
\begin{equation}{}
	\tilde S = \sqrt{\mathcal{M}} \cdot S.
\end{equation}
Due to the finite length of the TDS, the induced $\Delta \alpha$ can also lead to a significant change of the beta-function $\Delta \beta$ at the focusing element, which is relevant even for $d_l = 0$. The result is an overcompensation of the defocusing, or in other words underestimation of the bunch length. In order to determine the feasibility of using the compensation scheme discussed above, we have to consider the induced change in transverse normalized momentum along the TDS
\begin{equation}
	\Delta p_y^\prime = K\cdot \Delta s_0 + \frac{K^2}{\beta \gamma ^2} \cdot \frac{L}{6}\cdot y,
\end{equation}
which is just Eq.~\ref{eq:py_along_the_tds} devided by $p$. The induced position offset at the end of the TDS of length $L$ is then given by
\begin{equation}
	\begin{split}
		\Delta y &= \int K \cdot \left( \Delta s_0 + \frac{K}{\beta \gamma ^2}\cdot \frac{L}{6} \cdot y\right) \diff L\\
		&=  \underbrace{\frac{KL}{2} \cdot \Delta s_0}_{\text{Streaking}} + \underbrace{\frac{K^2}{\beta \gamma ^2} \cdot \frac{L^2}{24}\cdot y}_{\text{Defocusing}}.
	\end{split}
\end{equation}
In order to be able to use the compensation scheme, we have to demand that the second term is much smaller than the first one. Hence,
\begin{equation}
	\frac{\sigma_z}{\sigma_0} \overset{!}{\gg} \frac{KL}{12 \beta \gamma ^2},
\end{equation}
where $\Delta s_0$ and $y$ were replaced by $\sigma_z$ and $\sigma _0$ respectively. From this it is possible to define the streaking voltage criterion
\begin{equation}
	V \overset{!}{\ll} \frac{\sigma_z}{\sigma_0} \cdot \frac{12 \tilde p \beta \gamma ^2}{k L}
	\label{eq:voltage_criterion_compensation}
\end{equation}
for given initial beam and TDS parameters.

Since this criterion does not convey any information about the amount of overcompensation, nor the ideal value of $V$, it is necessary to conduct simulations. In these simulations one has to evaluate the relative error of the reconstruction with compensation $\sigma_{z,\text{rec}}$ compared to the theoretical resolution without bunch lengthening. i.e.
\begin{equation}
	\epsilon = \frac{\sigma_{z,\text{rec}}-\sqrt{\sigma^2_{z,\text{min}}+\sigma^2_{z,\text{0}}}}{\sqrt{\sigma^2_{z,\text{min}}+\sigma^2_{z,\text{0}}}}.
	\label{eq:relative_error}
\end{equation}
Then, one can define a threshold relative error $\epsilon_\text{thr}$ and determine a suitable streaking voltage based on Eq.~\ref{eq:voltage_criterion_compensation}. Note that this procedure of course only makes sense as long $|\epsilon _\text{thr}|$ is smaller than
\begin{equation}
	\epsilon_\text{dis} = \frac{\sqrt{\sigma_{z,\text{min}}^2 + \sigma_{z,0}^2 + \overline{\Delta\sigma}_{z,\text{dis}}^2}-\sqrt{\sigma^2_{z,\text{min}}+\sigma^2_{z,\text{0}}}}{\sqrt{\sigma^2_{z,\text{min}}+\sigma^2_{z,\text{0}}}},
\end{equation}
or in other words the compensation must not introduce a larger measurement error than the bunch lengthening itself.

\begin{figure}[htbp]
  \centering
  \includegraphics[width=\columnwidth]{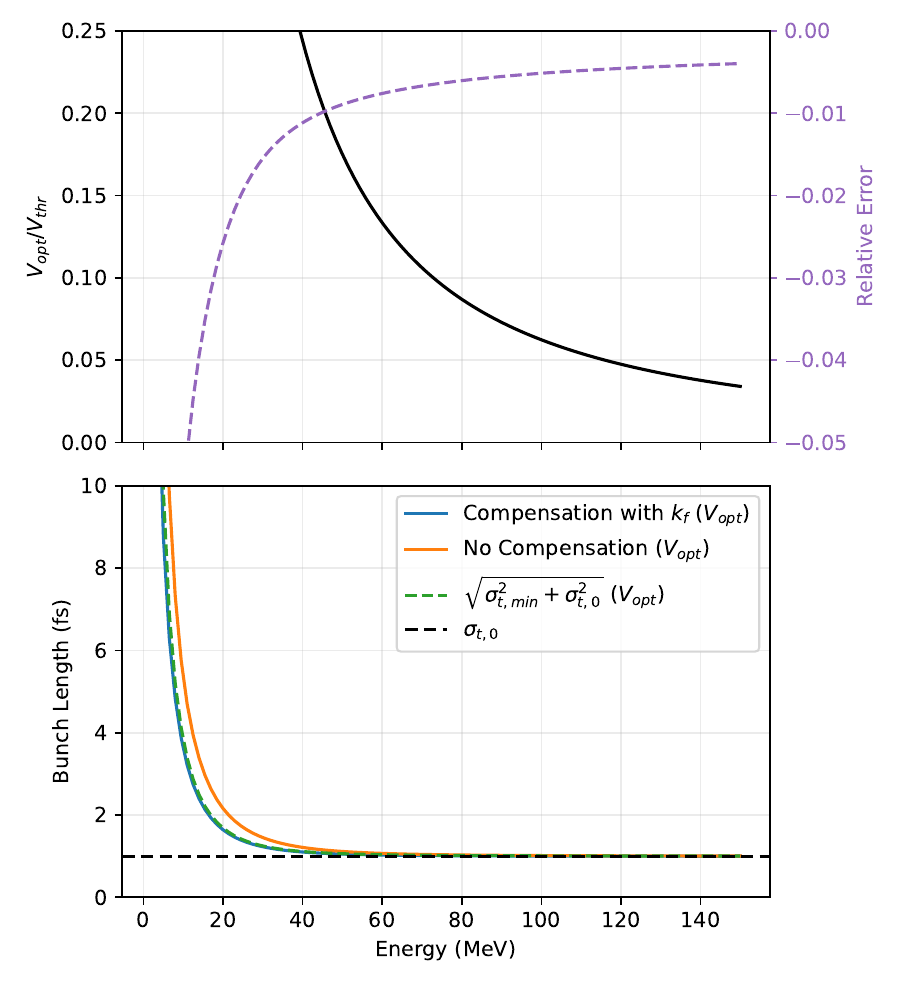}
  \caption{Tracking simulation results based on the same parameters as shown in Fig.~\ref{fig:POLARIX_example_tracking}, but including a compensation element of strength $k_f$. Top: Plot of the ratio $V_\text{opt}/V_\text{thr}$, where $V_\text{thr}$ is the right side of Eq.~\ref{eq:voltage_criterion_compensation}, vs the beam energy. The purple dashed line corresponds to $\epsilon$ (Eq.~\ref{eq:relative_error}). Bottom: Comparison of the reconstruction result using $V = V_\text{opt}$ for $k_f = 0$ (orange), $k_f \neq 0$ (blue) and the analytically calculated theoretical resolution (green dashed) vs beam energy.}
  \label{fig:POLARIX_example_tracking_compensation}
\end{figure}

An example simulation based on the same setup as described in the caption of Fig.~\ref{fig:POLARIX_example_tracking}, but only using $V = V_\text{opt}$, is shown in Fig.~\ref{fig:POLARIX_example_tracking_compensation}. It can be seen that the compensation brings the reconstruction result close to the theoretical resolution without bunch lengthening, while the always negative $\epsilon$ suggests a slight overcompensation on the percent level, as expected from the discussion above. At an exemplary beam energy of \SI{10}{\mega e\volt} it can be seen that the measured bunch length with compensation is \SI{3.6}{\femto \second}, wheras without compensation it is \SI{5.3}{\femto \second}. Employing the compensation at this particular energy hence results in a $\sim 1.5\times$ better resolution, compared to the already significantly improved resolution due to setting $V = V_\text{opt} = \SI{0.7}{\mega \volt}$ (at $V = V_\text{max} = \SI{20}{\mega \volt}$ the measured bunch length would be \SI{101.1}{\femto \second}). From the top plot of Fig.~\ref{fig:POLARIX_example_tracking_compensation} it can be extracted that for an allowed $\epsilon_\text{thr}$ of \SI{-1}{\percent} in this particular setup the streaking voltage must be smaller than $0.2 \cdot \frac{\sigma_z}{\sigma_0} \cdot \frac{12 \tilde p \beta \gamma ^2}{k L}$.
\subsection{Reduction of resolution due to energy spread}
Chromatic effects due to an incoming energy spread, but even more due to the induced energy spread, which is often significantly larger than the incoming energy spread, can impair the resolution of the TDS setup as well. While in case of a pure drift between TDS and screen the induced energy spread has no impact, a focusing element behind the TDS (see Appendix) will introduce additional chromatic aberrations. Ignored in the analytical treatment is also the dependence of the transverse kick of the TDS itself $y^\prime = \frac{V}{\tilde p} \sin(\phi) \approx K\zeta$.

A detailed discussion of chromatic effects is beyond the scope of this paper, it is however worth noting that the perturbed optics introduced above can be extended to incorporate chromatic focusing. It is thus a basis for the development of elaborated optical solutions alleviating also chromatic effects in TDS setups. The tracking simulations presented in this work include chromatic effects and show that in our case chromatic effects are insignificant.
\section{Other Resolution Limiting Factors}
Besides the fundamental TDS-related effects discussed above, other effects can impair the achievable effective resolution. For example, key performance characteristics of the experimental setup can complicate the bunch length measurement. Here, phase jitter ($\sigma_\phi$), a constant phase offset ($\Delta \phi$), as well as the achievable spatial resolution of the screen (\cite{Marx_2017} Eq.~8) should be mentioned. In addition, space charge forces can lead to defocusing and emittance growth, which in turn causes mismatched optics. It has furthermore been shown that in case of a strongly varying unperturbed bunch length during traversal of the TDS, the longitudinal focus needs to be either placed in the center of the structure, or such that the bunch length only changes linearly within the structure \cite{PhysRevLett.118.154802}.
\section{Conclusion}
We have derived an analytical estimate of the effective resolution of a TDS, which takes the induced energy spread and subsequent bunch lengthening inside of a finite length structure into account. From this, we define the optimal streaking voltage $V_\text{opt}$, which both depends on properties of the TDS, as well as the beam itself. At $V = V_\text{opt}$ the sweet spot between a resolution limit due to the resolving power of the TDS and the resolution reduction due to bunch lengthening is achieved. This concept is verified using matrix-based particle tracking simulations. Based on the technical parameters of the PolariX X-band TDS \cite{PhysRevAccelBeams.23.112001}, which is installed at the ARES linac at DESY, we show that within the range of frequently used beam energies at ARES (\SI{50}{\mega e\volt} - \SI{150}{\mega e\volt}) the maximum available streaking voltage of \SI{20}{\mega\volt} can indeed be too large and $V_\text{opt}$ needs to be employed.

In addition of discussing other resolution limiting effects, we furthermore study the notion of compensating the bunch lengthening induced defocusing by introducing a focusing element between the TDS and the screen. In case the induced change in beta-function is small enough, a compensation focusing strength $k_f$ can be calculated. Realistic setups with finite distance between the TDS exit and the focusing element can be employed, by taking the resulting de-magnification into account.

In conclusion, knowledge of the optimal streaking voltage is crucial for optimizing the performance of a TDS in low-$\gamma$ regimes. This is important for existing setups, but also for potential new designs. Conversely, it is important to realize that certain beams might not be possible to characterize with a given TDS setup. In specific cases, however, the compensation scheme discussed above can help, but needs careful preparation.
\begin{acknowledgments}
The authors acknowledge support from DESY (Hamburg, Germany), a member of the Helmholtz Association HGF. Furthermore, the authors thank M.~Kellermeier for thorough proofreading and discussions.
\end{acknowledgments}
\appendix*
\section{Optics examples for achieving optimal phase advance} \label{ap:optics_examples}
Assume a drift of length $l$ between TDS and screen and a focus located at $l_\text{f}$ in that drift. The unstreaked beta-function reads as
\begin{equation}
	\begin{split}
		\beta_{\text{u},0} &= \beta_\text{u,f} + \frac{l_\text{f}^2}{\beta_\text{u,f}},\\
		\beta_{\text{u,scr}} &= \beta_\text{u,f} + \frac{(l-l_\text{f})^2}{\beta_\text{u,f}},
	\end{split}
	\label{eq:beta_function_optics_example_focus_between_TDS_and_screen}
\end{equation}
where $\beta_\text{u,f}$ is the unstreaked beta-function at the focus. The phase advance follows from the definition
\begin{equation}
	\Delta \Psi = \int \frac{1}{\beta(s)} \diff s
\end{equation}
as
\begin{equation}
	\Delta \Psi = \arctan \left( \frac{l_\text{f}}{\beta_\text{u,f}} \right) + \arctan \left( \frac{l-l_\text{f}}{\beta_\text{u,f}} \right).
	\label{eq:phase_advance_optics_example}
\end{equation}
The condition $\Delta \Psi = \pi / 2$ can be fulfilled by various combinations of $l$ and $l_\text{f}$. For a given distance $l$ and beta-function in the TDS the system is however fully determined. Here, two example cases are discussed.

In case the focus is located very near to the screen, we can approximate $\beta_\text{u,f} \approx \beta_\text{u,scr}$ and $l_\text{f} \approx l$. Then, using the fact that $\arctan(0) = 0$, Eq.~\ref{eq:phase_advance_optics_example} yields
\begin{equation}
	\Delta \Psi \approx \arctan \left(\frac{l}{\beta_\text{u,scr}} \right).
\end{equation}
Furthermore, from Eq.~\ref{eq:beta_function_optics_example_focus_between_TDS_and_screen} we get
\begin{equation}
	l \approx \sqrt{\beta_{\text{u},0}\beta_\text{u,scr}-\beta_\text{u,scr}^2},
	\label{eq:optics_example_drift_length}
\end{equation}
which leads to
\begin{equation}
	\Delta \Psi \approx \arctan \left(\frac{\beta_{\text{u},0}}{\beta_\text{u,scr}}  - 1\right).
\end{equation}
Considering that $\lim_{x\to \infty} \arctan(x) = \pi/2$, in order to satisfy $\Delta \Psi = \pi/2$, it is hence required to ensure an as small as possible focus on the screen. Based on Eq.~\ref{eq:optics_example_drift_length}, this implies a short drift length between TDS and screen. In summary, the optical functions at the center of the TDS are given by
\begin{equation}
	\begin{split}
		\beta_{\text{u},0} &= \beta_\text{u,scr} + \frac{l^2}{\beta_\text{u,scr}}, \\
		\alpha_{\text{u},0} &= -\frac{\beta_{\text{u},0}^\prime}{2} = -\frac{l}{\beta_\text{u,scr}},\\
		\gamma_{\text{u},0} &= \frac{1+\alpha^2_{\text{u},0}}{\beta_\text{u,scr}}.
	\end{split}
\end{equation}
In thin-lens approximation these conditions can also be realized with a thin focusing lens directly after the TDS. In this case, only the initial beta-function is adjusted with the optics in front of the deflector, while $\alpha_{\text{u},0}$ is free and corrected with the focusing lens.

A second extreme case for the optics in the drift is realized by locating the focus directly behind the TDS and thus $\beta_\text{u,f} \approx \beta_{\text{u},0}$ and $\alpha_{\text{u},0} = 0$. The beta-function is then given by
\begin{equation}
	\beta_\text{u,scr} \approx \frac{l^2}{\beta_{\text{u},0}}
\end{equation}
for $l \gg \beta_{\text{u},0}$. The phase advance develops as 
\begin{equation}
	\Delta \Psi \approx \arctan \left(\frac{l}{\beta_{\text{u},0}} \right).
\end{equation}
With this solution $l$ therefore needs to be maximized in order to fulfill $\Delta \Psi = \pi / 2$.

An intermediate case would be $l_\text{f} = l/2$. This means $\beta_\text{u,scr} = \beta_{\text{u},0}$ and the phase advance is given by
\begin{equation}
	\Delta \Psi = 2 \cdot \arctan \left( \frac{l_\text{f}}{\beta_\text{u,f}}\right).
\end{equation}
Since $2\cdot \arctan(1) = \pi/2$, this yields $l_\text{f} = \beta_\text{u,f}$, $l = 2\beta_\text{u,f}$, and subsequently using Eq.~\ref{eq:beta_function_optics_example_focus_between_TDS_and_screen} $l = \beta_{\text{u},0}$, in order to fulfill $\Delta \Psi = \pi / 2$.
\bibliography{TDSEffectiveResolution}
\end{document}